\documentclass{svjour2}
\smartqed
\usepackage{graphicx}
\usepackage{latexsym}
\usepackage{amsmath}
\usepackage{amsfonts}
\usepackage{amssymb}
\usepackage{epstopdf}
\usepackage{mathptmx}
\usepackage{url} 
\usepackage{cite}

\journalname{Foundations of Physics}
\begin{document}
\bibliographystyle{spphys} 
\title{Quantum statistics as geometry: Conflict, Mechanism, Interpretation, and Implication}
\subtitle{}
\titlerunning{Quantum statistics as geometry}  
\author{Daniel C. Galehouse}
\institute{D. Galehouse \at 
					University of Akron \\
					Tel.: 330-658-3556\\ 
					\email[dcg@uakron.edu] \\
					\emph{15764 Galehouse Rd. Doylestown, Ohio 44230} \\
					homepage: [\url{http://gozips.uakron.edu/~dcg}] \\}
\date{Recieved: date Accepted: date Today: \today}
\maketitle

\begin{abstract}

The conflict between the determinism of geometry in general relativity and the essential statistics of quantum mechanics blocks the development of a unified theory.  Electromagnetic radiation is essential to both fields and supplies a common meeting ground.  It is proposed that a suitable mechanism to resolve these differences can be based on the use of a time-symmetric treatment for the radiation.  Advanced fields of the absorber can be interpreted to supply the random character of spontaneous emission.  This allows the statistics of the Born rule to come from the spontaneous emission that occurs during a physical measurement.   When the absorber is included, quantum mechanics is completely deterministic. It is suggested that the peculiar properties of kaons may be induced by the advanced effects of the neutrino field.  Schr\"odinger's cat loses its enigmatic personality and the identification of mental processes as an essential component of a measurement is no longer needed. 
 
\keywords{geometry \and quantum mechanics \and radiation \and determinism \and general relativity}
\PACS{01.30.Cc \and 02.40.Ky \and 03.65.Ta \and 04.50.-h \and 14.40.Aq }
\subclass{ 51Fxx \and 51PO5 \and 81Q70 \and 81S99 \and 83E15}
\end{abstract}

\section{Conflict \label{sec:intro}}

The conceptual clash between the statistical nature of quantum mechanics and the deterministic character of gravity has persisted since the early days of modern physics.  Here, a mechanistic description is offered that can account for fundamental statistics in a fully geometrical theory.  It is a natural construction for a geometry based quantum theory and allows other developments to progress.  Larger structures involving more dimensions can extend the range to charge, spin, and weak or strong interactions.~\cite{fgqp,qgtis,pepscs}

In these geometrical theories, particles are described exclusively by wave functions having the conventional quantum properties.  All interactions are mediated by conformal transformations that are applied to a geometry suited to the properties of the individual particle.   Quantum motion comes from the time development of space-time distortions.  There are no classical point objects.  Particle dynamics follows from an invariant wave equation.   Any Hilbert space is introduced as a means of calculation and is subsidiary to the field equation.  A wave function expansion may be useful, but it is only the sum total wave-function that is identified with any part of the geometrical system.   The evolution of the system of wave fields follows a Dirac-like equation and is completely deterministic. This article discusses the foundational issues that are relevant to this approach.  

Without classical particles it is difficult to formulate Laplace's notion of determinism.  Wave-particles may extend to the edges of space, and perhaps the whole universe.  Such an object, subject to interaction, but ultimately fixed in space-time, is to be accepted as the target of Laplace's concept.   The assertion of determinism, that events at a later time are entirely calculable from those at an earlier time, is difficult to apply because the particles are never completely localized.  All events, past and future, are set in the geometry.    The individual solution of such a system of equations can have no formal statistics.  Perhaps one could denote it as 'mechanistic' or 'absolute' determinism.  Time reversal is an exact fundamental symmetry.  The notions of causality and determinism that are used in conventional measurement theory must be developed from the properties of the electromagnetic interactions.  

\section{Mechanism \label{sec:mech}}

\begin{figure}[ht]
\includegraphics[height=2.cm ]{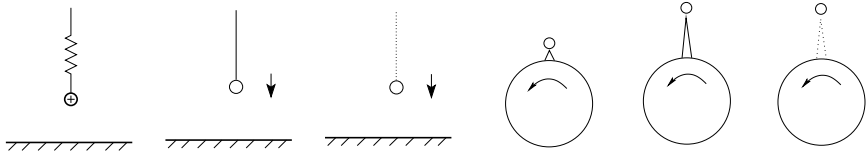}
\caption{ If a charged particle oscillates on a spring, radiation is emitted.  The rate of radiation is calculated from the acceleration relative to the local inertial frame.  If it hangs still, no radiation is emitted.  But in general relativity it is not that simple.  The stationary particle in a gravitational field is no longer in an inertial frame, it has a constant upward acceleration but does not radiate.  If the particle is allowed to fall, it is then in an inertial frame even though it accelerates downward and does radiate.  More interestingly, if the particle is put, at a fixed height on the top of a tower that rotates with the earth, it will radiate from the rotation but not from the acceleration of gravity.  If now, the tower is made taller, the radiation increases.  This radiation persists even if the height is increased to the geo-synchronous point.  The motion is now inertial but it must still radiate.  If the tower is removed, the radiation is unchanged (to lowest order) but the energy comes from the gravitational potential of the mass above the earth rather than the rotational kinetic energy of the earth itself.}
\label{fig:1}
\end{figure}

Figure~\ref{fig:1} offers some examples from radiation theory that are intended to illustrate the difficulties of describing radiation in a general relativistic setting.  The usual simple concept that uses the acceleration relative to a local inertial frame is to be found insufficient .  A difficult general case is that of a charged mass coasting among gravitating bodies.  There is no simple definition of acceleration, no characteristic scale of distance, no assured velocity limit, no characteristic wavelength, and no guarantee of a simple expansion of the radiation field.  These issues all defy the usual interpretation.  It should be noted, that because of diffraction, a quantum particle cannot be assigned a constant acceleration relative to any inertial frame.  The usual concept of radiation fails when acceleration is not absolute.   A proper theory must handle all cases as well as the most general possibility of additional non-inertial forces (such as weak or strong interactions).  

More difficult is the calculation of radiation reaction forces. These extract energy from the emitter.  Usually, they are derived after the radiation itself is known. That these forces are secondary to the interaction cannot be held for a geometrical theory.  Consider two closely spaced emitters transmitting in phase.  The power of each alone depends on the respective driving current squared.  For both emitters together, the total power is proportional to the square of the sum of the currents.  (The electric fields are in phase and interfere constructively.) The total emission is greater than the sum of the parts. The additional power must be supplied by the driving system.  The driving voltage increases.  A simple calculation shows that this increased reaction is equivalent to the energy required to move the current of each antenna through the reaction field of the other.   This behavior has been known in atomic spectroscopy since the early days.  In conventional quantum mechanics, an alternative construction is used which models such a process as the transfer of energy into or out of an abstract vacuum field.  

These issues, in a general relativistic system, are best handled by the mathematical device of the covariant two-point tensor.  These green's functions act between separated positions in space-time.  The two point tensor has indicies that transform covariantly at either end.  Thus, a radiative interaction can be described in a systematic way without depending on inertial frames. In this context, a time-symmetric formulation may be used.~\cite{davies1,davies2,davies3,leiter6,leiter1}   Problems with frame dependence are resolved and the radiative forces are described consistently.  All forces have an ontological geometrical origin.  Quantum radiation forces, once interpreted as vacuum effects, are here due to the advanced fields of other particles. 

A proper analysis would require a covariant form of quantum electrodynamics.  A complete construction is not yet available.  The geometrical theories are exact and non-perturbative as written.  These may eventually lead to a structure in which the conventional perturbative expansion appears as a limiting case.  It is a difficult problem because any attempt to formulate a non-perturbative quantum electrodynamics requires the use of curvilinear fields. Fortunately, conventional quantum electrodynamics is sufficient for the foundational issues considered here. 

\section{Interpretation \label{sec:int}}
\begin{figure}[ht]
\includegraphics[height=1.85cm ]{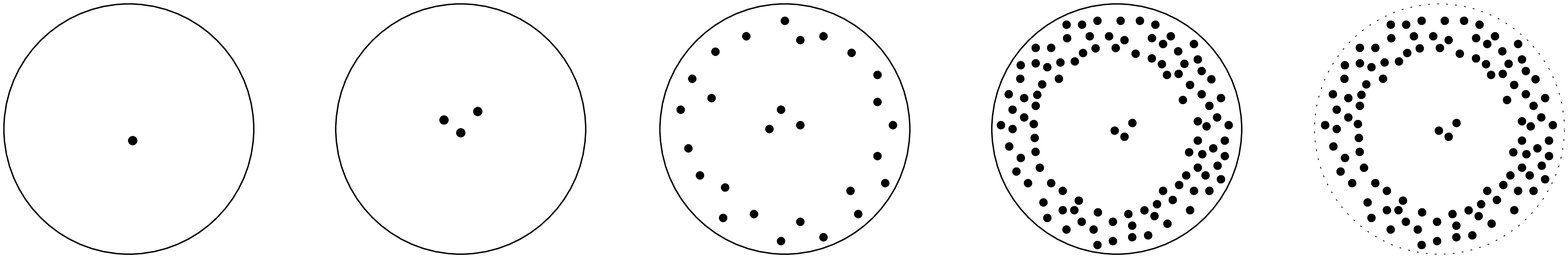}
\caption{Starting from a perfectly reflecting empty cavity, particles are added, one by one, to build up an experimental apparatus. Perhaps $k$ of them, will be arranged to make the experiment itself.  In addition, $n$ particles will make up an absorber. These are placed around the outside and exchange photons with the experiment.  As $n$ becomes very large, the exterior matter is sufficient to absorb or emit all of the photons that may be required by the experiment.  For sufficiently large $n$, the mirror may be removed.}
\label{fig:3}
\end{figure}

As shown in figure~\ref{fig:3}, a cavity is built up to contain particles that evolve deterministically.  There are no statistics since no formal measurements are being made.  Photons will be exchanged  between $k$ particles making up the experiments and the $n$ particles outside.  The outer particles contribute to the random character of the spontaneous emission.  For large $n$, normal radiative behavior is obtained.  These absorbing particles participate in the transitions in lieu of the vacuum system.  One concludes that if the absorbing particles are included in the description, the quantum mechanics is completely deterministic.  This is the main result: The statistics of spontaneous emission are compatible with a mechanistic geometry.

Spontaneous emission occurs as the advanced fields come in from the future. The result is a contribution to random behavior having no identified cause. The questions are: 'What causes spontaneous decay?' and 'What are the unpredictable effects of advanced radiation?'. The answer is that they are aspects of the same thing.  The advanced interactions initiate the electromagnetic decay and account for the randomness in the spontaneity of emission.  In a geometrical theory, the non-local character of the interaction derives from the originating conformal transformations.  An advanced or retarded interaction is only a mathematical representation of part of the whole interaction. (As noted below, the non-random effects of the advanced fields may also contribute to an entanglement.)

\section{Implication  \label{sec:app}}

\begin{figure}[ht]
\includegraphics[height=2.5cm ]{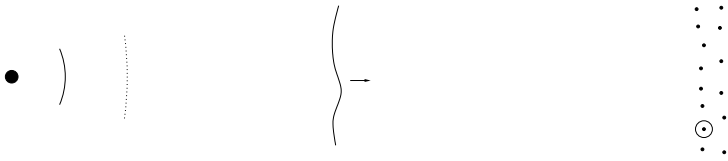}
\caption{Beta particles are emitted from the nucleus and travel, as a wave, to a detector.  They cascade to final states on the screen.  A collection of bare protons are used as a simplified model for the screen.   A probability distribution is built up from multiple observations of particle arrivals.  Each electron comes to rest, as a wave, on one of the detector sites.  Radiation is emitted during the time of capture.}
\label{fig:4}
\end{figure}

Given this understanding of spontaneous emission, it is possible to give an explanation for the fundamental statistics of quantum particles as they are observed in a diffraction experiment.  The proposition is that all such measurement statistics derive ultimately from spontaneous emission that occurs during the measurement itself.  Born's experiment,~\cite{bornpd}, as illustrated in figure~\ref{fig:4}, shows a nuclear decay and the associated detection as an emitted electron collides with the screen.  Born's arguments assign the square of the absolute value of the wave function to the probability density.  Here, the individual detection events are effectively selected by the statistics of spontaneous emission.  
An emitted electron travels to the screen, and eventually comes to rest as a bound wave function on an active detector site.  Along the way, it will emit radiation as it cascades through one or more intermediate states.  The forces of radiative reaction act to change the quantum state of the electron as it progresses. Energy is emitted. If it ends up on an upper proton, the radiation is different, in detail, than if it ends up on a lower proton.  This radiation is sometimes called cascade radiation or, at higher energies, bremstralung.  If the emitted electromagnetic fields are time-reversed, the electron will be removed from the proton and be returned to the nucleus.   The information in the bremstralung contains all details of how the electron made its transition from the initial state to the final position.  The process appears to be statistical if the information carried away by the radiation is lost to the observer.  The 'choice' of a particular proton will appear random and will depend on the outcome of the various uncontrolled competitive spontaneous emission processes.  

Because the  electromagnetic transitions, are proportional to the 'matrix element squared' or 'transition probability', the chance of any given final position (say on a particular individual proton) is proportional to the square of the initial wave function.   Of course a good measurement requires that the transition to each final position be equivalent.  The observed distribution of final states thus follows from the original probability wave.  This bias, affecting the spontaneous rate, is carried through the cascade to the final distribution.

A practical detector gives an accurate probability down to the size of the final electron state.  No classical point events are used, the measurement is adequately modeled with wave functions.   A general quantum measurement is an abstraction of such a quantum-electromagnetic interaction.   (This interpretation may be extended to other force fields.)  In this way, the statistics of particle observations appear.  The real statistical effects originate in the absorber and the underlying theory remains deterministic.

\begin{figure}[ht]
\includegraphics[height=2.5cm ]{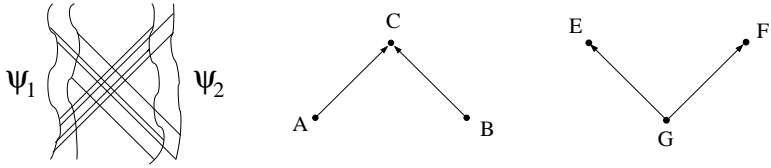}
\caption{An entanglement is an ongoing interaction, forward and backward, between the collected elements of two wave-functions.  The states change continuously and may undergo binding or scattering.  Some transitions require the absorption or emission of two photons simultaneously.  Correlations are enforced.  A double absorption at C may be satisfied by the simultaneous arrival of two photons from A and B.  A correlated pair emission from a particle at G may be identified by coincidence timing of events at E and F. The enigma is to understand how any effect could get from B to A since they are space-like separated.   Retrodiction from C to B is not allowed classically.}
\label{fig:5}
\end{figure}

In figure~\ref{fig:5}, quantum states which interact non-linearly with the electromagnetic field force correlations or entanglements between photons.  As part of this process, the advanced and retarded interactions must also replace the space-like interactions that are used in classical physics but which are not accepted in relativity.  The advanced part of the interaction contributes to the correlations of an entangled state. A spin or position correlation may be created.  Radiative reaction terms are essential, insuring that a measurement of one of the correlated particles includes a measurement of the other.

\begin{figure}[ht]
\includegraphics[height=2.45cm ]{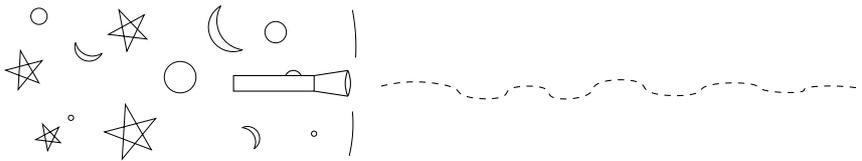}
\caption{A flashlight points outward at the edge of the universe.  If there are no particles to absorb radiation can energy be emitted?}
\label{fig:6}
\end{figure}

The problem of retrodiction can be explored more carefully. The immediate question, as posed in figure~\ref{fig:6}, is whether electromagnetic energy will be emitted without any absorber present.~\cite[See footnote 23, p455]{feynmanmf}. If the light is allowed to continue indefinitely it will permanently  violate mass, energy, and momentum conservation. The heat energy of the filament will be lost to eternity.  Alternatively, If light cannot be emitted, the filament will have an unexpectedly higher temperature.   At the present time, there are no identified observations of any such effect; however, in a later cosmological epoch, the density of distant particles will be less and a reduction in absorption should occur.~\footnote{This issue is also related to the information paradox in black hole theory.}

\begin{figure}[ht]
\includegraphics[height=3.2cm ]{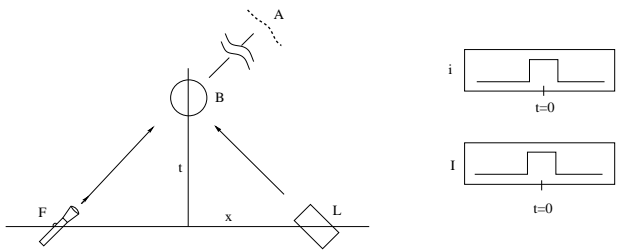}
\caption{A flashlight, F, points through a transparent cell, B, at a region of the cosmos, A, having deficient absorption.  The filament temperature is higher because the light emission is inhibited.  The transparent cell contains a gas that can be made opaque by an optical signal from an auxiliary source, L.  A burst of radiation will opacify this intermediate absorber.  At this time, the flashlight emission will increase and the filament temperature will decrease.  The resistance decreases and the filament current increases coincident with the emission of the switch pulse. In the local system as shown, the signal from point L, at t=0, appears at the space like separated point F, at t=0.}
\label{fig:7}
\end{figure}

For geometrical theories, ontology requires that no photons be emitted without being absorbed.  If this true, and if there are absorber limitations to be found, space-like transmission of signals may be possible.  The space time diagram is shown in figure~\ref{fig:7}.  A switchable absorber is placed in front of the flashlight.    A burst of radiation switches on the absorber which retrodicts to the flashlight.  The process is in accord with relativity if the reverse signals travel on null lines.

Observational tests are difficult. They have not been found possible for photons but other null fields may show an effect.  At present, tests using gravitation are open.  In so far as neutrinos are fundamentally null, they may supply such an example of retrodiction, especially since the cosmological neutrino opacity is well below that of the photon. The studies of kaon decay are extensive and not completely conclusive but available data may already show such an effect.~\cite{rpp2012}  Elementary discussions of kaons are available.~\cite{fitchcpas,cronincpsa,kabircppuz}  The approach of this paper forces a modification of the vacuum and requires corrections to the Weisskopf-Wigner formalism,~\cite{weisskopfw}, including its use for neutrino emission.  Many questions are still open; see for example~\cite{didomhur,aronsoned}.   

There are some elementary considerations expected for the effects on kaon decay.  The $K_0$ and its antiparticle, the $\overline{K_0}$ are putative charge-parity conjugates.  In a vacuum that is explicitly neutrino-antineutrino unsymmetrical, there may be a preference for some decays over others.  Differences in resonance width or lifetimes may be induced. The decay and time evolution of the particle may then show deviations from expectations based on exact CP invariance.  Forces that violate CP invariance are brought in by the advanced fields.   The preference for absorption of anti-neutrinos by protons and neutrinos by neutrons should leave an imbalance that depends on the neutron to proton ratio within the background region of beam absorption.   An apparent violation of CP invariance is possible even if the fundamental physical dynamics is invariant.  Such processes may confound the currently accepted  interpretations of kaon decay.  In time, the sensitivities of bench experiments that do not emit neutrinos to infinity may deny the existence of explicit CP breaking interactions.  It is an important question for geometrical theories as no structural violation of CP invariance is expected.   

Accordingly, different absorbing backgrounds may affect decay rates.  Sensitivities are hard to predict, but known neutron to proton ratios do vary as much as ten percent for short distances.  The many known observations of kaon decay may be shown, with further analysis, to have a background dependence.  One might look at how kaons decay toward the oceans, or the crust, or the core of the earth.  There may be tests involving the galaxy or solar system.  The issue is important for quantum foundations because, if the advanced effects can be proven, the absorber must then contribute explicitly to the statistics of spontaneous emission.  

These constructions abrogate standard measurement theory. VonNeumann proposed a separate type of evolution for quantum measurements.   A formal process, existing in the classical domain, was to be associated with the mental recognition of an observation.   In the present era, our understanding of complex things, living and non-living is more complete, and the separation between them is no longer distinct.   The identification of a separate mental process as part of a measurement is not as compelling as it once was.  The combined classification of living things, such as cats, with inanimate objects, such as computers seems more reasonable now.  We also understand that neither our thought processes nor the machinations of computers are to be modeled classically.  For a complex device, elementary statistical variations in the output can be due to deficiencies of the processing mechanism or errors in the incoming information.  These appear to us as deterministic.   In addition, other variations in output come from the retrodiction of the advanced fields and are identified as true fundamental noise.  This noise is equivalent to spontaneous emission and cannot be eliminated.  It appears to us as truly random, but for the universe as a whole may be considered absolutely deterministic.

Also now, the cat paradox~\cite{schrodinger8} is elementary.  The radioactive decay will require a free absorber for the emitted neutrino.  This neutrino will take with it the angular momentum, that must be removed. The electron will set off the counter that will release the poison and and kill the cat.  The process depends on the absorber; the randomness of the nuclear decay need no longer be considered fundamental. The cat either dies or lives.  There are no superpositions.  You may open the box at any time to see what has happened.

These are the implications of a fundamental geometry for the quantum.  A mechanistic approach leads to the universal use of time-symmetry in all interactions.  Spontaneous emission is identified with the advanced fields, and the origin of the statistics of the probability density are from the electromagnetic cascade.  Experimental tests may confirm or deny these presumptions.  Is it God or Mother nature that provides the spontaneous behavior of person, cat, or atom? 

\section{Acknowledgement \label{sec:ack}}

This article is based on a talk given at the Conference "Quantum Theory from Problems to Advances", June 9-12, 2014, Linnaeus University, Sweden.

\bibliography{qdet}

\end{document}